\documentclass[journal,one column,12pt]{IEEEtran}
\usepackage{amsfonts}
\usepackage{amssymb}
\usepackage{graphicx}
\usepackage{amsmath}
\usepackage{epsfig}
\usepackage{setspace}
\usepackage{color}
\newtheorem{thm}{Theorem}
\newtheorem{definition}{Definition}
\newtheorem{lemma}{Lemma}

\newcommand{\mv}[1]{\mbox{\boldmath{$ #1 $}}}

\begin{document}

\title{Outage Capacity and Optimal Transmission \\ for Dying Channels
\thanks{Part of the work was presented at GLOBECOM'08 and ICC'09,  and is supported in part by
the National Science Foundation under Grant CNS-0721935, and by the
Department of Defense under Grant HDTRA-07-1-0037.}}

\author{Meng Zeng, Rui Zhang, and Shuguang Cui
\thanks{Meng Zeng and Shuguang Cui are with the Department of Electrical and Computer
Engineering, Texas A\&M University, College Station, TX, 77843. Emails:
\{zengm321, cui\}@tamu.edu.}
\thanks{Rui Zhang is with the Institute for Infocomm Research, A*STAR,
Singapore. Email: rzhang@i2r.a-star.edu.sg.}}

\maketitle
\begin{abstract}
In wireless  networks, communication links may be subject to random
fatal impacts: for example, sensor networks under sudden power losses
or cognitive radio networks with unpredictable primary user spectrum
occupancy. Under such circumstances, it is critical to quantify how
fast and reliably the information can be collected over attacked links.
For a single point-to-point channel subject to a random attack, named
as a \emph{dying channel}, we model it as a block-fading (BF) channel
with a finite and random delay constraint. First, we define the outage
capacity as the performance measure, followed by studying the optimal
coding length $K$ such that the outage probability is minimized when
uniform power allocation is assumed. For a given rate target and a
coding length $K$, we then minimize the outage probability over the
power allocation vector $\mv{P}_{K}$, and show that this optimization
problem can be cast into a convex optimization problem under some
conditions. The optimal solutions for several special cases are
discussed.

Furthermore, we extend the single point-to-point dying channel result
to the parallel multi-channel case where each sub-channel is a dying
channel,  and investigate the corresponding asymptotic behavior of the
overall outage probability with two different attack models: the
independent-attack case and the $m$-dependent-attack case. It can be
shown that the overall outage probability diminishes to zero for both
cases as the number of sub-channels increases if the \emph{rate per
unit cost} is less than a certain threshold. The outage exponents are
also studied to reveal how fast the outage probability improves over
the number of sub-channels.

\end{abstract}
\newpage
\section{Introduction}
Information-theoretic limits of fading channels have been thoroughly
studied in the literature and to date many important results are known
(see \cite{Shamai98} and references therein). Generally speaking, if
the transmission delay is not of concern, the classic Shannon capacity
for a deterministic additive white Gaussian noise (AWGN) channel can be
extended to the ergodic capacity for a fading AWGN channel, which is
achievable by a random Gaussian codebook with infinite-length codewords
spanning over many fading blocks such that the randomness induced by
fading can be averaged out \cite{Goldsmith97}\cite{Shamai99}. With the
transmitter and receiver channel state information (CSI) perfectly
known, the adaptive power allocation serves as an effective method to
increase the ergodic capacity. This allocation has the well-known
``water-filling'' structure \cite{Goldsmith97}, where power is
allocated over the channel state space. With such an allocation scheme,
a user transmits at high power when the channel is good and at low or
zero power when the channel is poor. When the CSI is only known at the
receiver, the capacity is achievable with special ``single-codebook,
constant-power'' schemes \cite{Caire99}.

The validity of the ergodic capacity is  based on the fundamental
assumption that the delay limit is infinite. However, many wireless
communication applications have certain delay constraints, which limit
the practical codeword length to be finite. Thus, the ergodic capacity
is no longer a meaningful performance measure.  Such situations give
rise to the notions of \emph{outage capacity}, \emph{delay-limited
capacity}, and \emph{average capacity} \cite{Berry00}\cite{Yeh06}, each
of which provides a more meaningful performance measure than the
ergodic capacity. In particular, there usually exists a
capacity-versus-outage tradeoff for transmissions over fading channels
with finite delay constraints \cite{Shamai94}, where an outage event
occurs when the ``instantaneous'' mutual information of the fading
channel falls below the transmitted code rate, and a higher target rate
results in a larger outage probability. The maximum transmit rate that
can be reliably communicated under some prescribed transmit power
budget and outage probability constraint is known as the outage
capacity.  In the extreme case of requiring zero outage probability,
the outage capacity then becomes the zero-outage or delay-limited
capacity \cite{Tse98}. To study the delay-limited system, the authors
in \cite{Berry00} adopt a $K$-block block-fading (BF) AWGN channel
model, where $K$ indicates the constraint on transmission delay or the
maximum codeword length in blocks. Such a channel model is briefly
described as follows. Suppose a codeword is required to transmit within
$KB$ symbols, with the integer $K$ being the number of blocks spanned
by a codeword, which is also referred to as the interleaving depth (we
call it coding length to emphasize how many blocks over which a
codeword spans); it is also a measure of the overall transmission
delay. The parameter $B$ is the number of channel uses in each block,
which is called block length. A codeword of length $KB$ is also
referred to as a frame, where the fading gain within each block remains
the same (over $B$ symbols) and changes independently from block to
block. The number of channel uses $B$ in each block is assumed to be
large enough for reliable communication, but still small compared to
the channel coherence time. If the CSI for each $K$-block transmission
is known non-causally at the transmitter, transmit power control can
significantly improve the outage capacity of the $K$-block BF channel
\cite{Caire99}. When the CSI can be only revealed to the transmitter in
a causal manner, a dynamic programming algorithm is developed to
achieve the outage capacity of the $K$-block BF channel in
\cite{Negi02}.

In the above existing works, the delay limit is either infinite or
finite but deterministic. However, there are indeed some practical
scenarios where the delay constraint is both finite and random. For
example, in a wireless sensor network operating in a hostile
environment, sensors may die due to sudden physical attacks such as
fire or power losses. Another example may be a cognitive radio network
with opportunistic spectrum sharing between the secondary and primary
users, where an active secondary link can be corrupted unpredictably
when the channel is reoccupied by a primary transmission. How fast and
reliably can a piece of information be transmitted over such a channel?
This question motivates us to formally define the maximum achievable
information rate over a channel with a random and finite delay
constraint, named as a \emph{dying channel}. This type of dying
channels has never been thoroughly studied in the traditional
information theory, and important theorems are missing to address the
fundamental capacity limits. In this paper, we start investigating such
channels by focusing on a point-to-point dying link and model it by a
$K$-block BF channel subject to a fatal attack that may happen at a
random moment within any of the $K$ transmission blocks, or may not
happen at all over $K$ blocks. Note that the delay limit in the case of
a dying channel is a random variable due to the random attack, instead
of being deterministically equal to $K$ as in a traditional
delay-limited BF channel. Since the successfully transmitted number of
blocks is random and up to $K$, a dying channel is delay-limited and
hence non-ergodic in nature. Thus its information-theoretic limit can
be measured by the outage capacity. It is well known that coding over
only one block of a fading channel may lead to a poor performance due
to the lack of diversity. However, when we code over multiple blocks to
achieve more diversity in a dying channel, we must bear the larger
possibility that the random attack happens in the middle of the
transmission and renders the rest of the codeword useless. Therefore,
it is neither wise to span a codeword over too many blocks nor just
over one block. We need to consider the tradeoff between the potential
diversity and the attack avoidance for the selection of the codeword
length over such a dying channel. In other words, given a distribution
of the random attack, we need to seek an optimal $K$ that ``matches''
the number of surviving blocks in a probabilistic sense such that the
achievable diversity is maximized and the outage probability is
minimized.

In a system with multiple parallel sub-channels (e.g., in a OFDM-based
system), each sub-channel may be under a potential random attack. In
such a scenario, we are interested in the overall system outage
probability and how the outage probability behaves as the number of
sub-channels increases. This leads us to examine the asymptotic outage
behavior for the case of a parallel dying channel. We will consider two
models of random attacks over the sub-channels: 1) the case of
independent random attacks, where the attacks across the sub-channels
are independently and identically distributed (i.i.d.) ; and 2) the
case of $m$-dependent random attacks, where the attacks over $m$
adjacent sub-channels are correlated and the attacks on sub-channels
that are $m$-sub-channel away from each other are independent.

In the following, we briefly summarize the main results in this
paper:
\begin{enumerate}
   \item  We introduce the notion of a dying channel and formally define its outage capacity.
    Suppose we code over $K$ blocks, and the number of surviving
    blocks is random and up to $K$. An outage occurs if the total mutual information over the
    surviving blocks normalized by $K$ is less than a predefined rate
    $R$. Correspondingly, the outage capacity is the largest rate that
    satisfies an outage probability requirement.
    \item  We study the optimal coding length $K$ that ``matches'' the attack time in a
    probabilistic sense such that the outage probability is minimized when uniform power
    allocation is assumed. We then investigate the optimal power allocation over these $K$
    blocks, where we obtain the general properties for the optimal
    power vector $\mv{P}_K$. We find that, for some cases, the optimization problem over
    $\mv{P}_K$ can be cast into a convex problem.
 \item  We further extend the single dying channel result to the parallel dying
    channel case where each sub-channel is an individual dying channel. In this case,
    we investigate the outage behavior with two different random attack models: the independent-attack case and the
    $m$-dependent-attack case. Specifically, we characterize the asymptotic behavior of the outage
    probabilities for the above two cases with a given target rate. By the central limit theorems for
    independent and $m$-dependent sequences, we show that the outage
    probability diminishes to zero for both cases as the number of
    sub-channels increases if the target rate per unit cost is below a
    threshold. The outage exponents for both cases are studied to reveal how fast the outage probability
    improves.
\end{enumerate}

The rest of this paper is organized as follows. Section \ref{sec:system
model} presents the system model for a single dying channel, as well as
the definition of the corresponding outage capacity. In Section
\ref{sec:uniform power}, we study the optimal coding length by
considering uniform power allocation and derive the lower and upper
bounds of the outage probability. Moreover, we obtain the closed-form
expression of outage probability for the high signal-to-noise ratio
(SNR) Rayleigh fading case. In Section \ref{sec: joint opt}, we
optimize over the power vector to minimize the outage probability. In
Section \ref{sec: parallel }, we extend the single dying channel model
to the parallel dying channel case. In particular, we examine the
corresponding asymptotic outage probability with two setups: the
independent-attack case and the $m$-dependent-attack case, in Sections
\ref{sec: indp case} and \ref{sec: m-dep case}, respectively. In
Section \ref{sec: outage exponent}, the outage exponents for both cases
are examined to reveal how fast the outage probability improves.
Section \ref{sec:conclusion} concludes the paper.

\emph{Notation}: we define the notations used throughout this paper as
follows.
\begin{itemize}
  \item $\mathbb{R}$ indicates the set of real numbers, $\mathbb{R}_{+}$ is the
  set of nonnegative real numbers, and $\mathbb{R}_{+}^{N}$ is the set of
  $N$-dimensional nonnegative real vectors.
  \item The error function: $\textrm{erf}(x)=\frac{2}{\sqrt{\pi}}\int_{0}^{x}e^{-t^2} dt$.
  \item The normalized cumulative normal distribution function:
  $\Phi(x)=\frac{1}{\sqrt{2\pi}}\int_{-\infty}^{x}e^{-\frac{t^2}{2}}dt$.
  \item The $Q$-function: $Q(x)=\frac{1}{\sqrt{2\pi}}\int_{x}^{\infty}e^{-\frac{t^2}{2}}dt$.
  \item $\log(x)$ is the natural logarithm.
  \item $\lceil \cdot \rceil$ is the ceiling operator and $\lfloor \cdot \rfloor$ is the flooring
  operator.
\end{itemize}

\section{Outage Capacity Definition of a Single Dying Channel}\label{sec:system model}
We consider a point-to-point delay-limited fading channel subject to a
random fatal attack, while the exact timing of the attack is unknown to
neither the transmitter nor the receiver. Only the distribution of the
random attack time is known to both the transmitter and the receiver.
We further assume that there is no channel state information at the
transmitter (CSIT) while there is perfect channel state information at
the receiver (CSIR). The transmitter transmits a codeword over $K$
blocks within the delay constraint; and when the fatal attack occurs,
the communication link is cut off immediately with the current and rest
of the blocks lost. We build our model of such a dying link based on
the $K$-block BF-AWGN channel \cite{Berry00}, which is described as
follows.

Let $\mathbf{x}$, $\mathbf{y}$, and $\mathbf{z}$ be vectors in
$\mathbb{R}^{KB}$ representing the channel input, output, and noise
sequences, respectively, where $\mathbf{z}$ is the Gaussian random
vector with zero mean and covariance matrix $\sigma^2 \mathbf{I}_{KB}$.
Rearrange the components of $\mathbf{x}$, $\mathbf{y}$, and
$\mathbf{z}$ as $K\times B$ matrices, denoted as $\mathbf{X}$,
$\mathbf{Y}$, and $\mathbf{Z}$, respectively (Each row is associated
with $B$ symbols from a particular block.). A codeword with length $KB$
spans $K$ blocks and the input-output relation  over the channel can be
written as follows:
\begin{equation*}
    \mathbf{Y}=\mathbf{A}\mathbf{X}+\mathbf{Z},
\end{equation*}
where $\mathbf{A}=\textrm{diag}(|h_1|, \cdots, |h_K|)$ is a $K\times K$
matrix with the diagonal elements being the fading amplitudes. Let
$\hat{\mathbf{X}}_i$ be the $i$-th column of $\mathbf{X}$ for $i\in
\{1,\cdots, B\}$. Similarly, let $\hat{\mathbf{Y}}_i$ and
$\hat{\mathbf{Z}}_i$ be the $i$-th columns of $\mathbf{Y}$ and
$\mathbf{Z}$, respectively. These are related as:
\begin{equation}\label{eq: parallel ch model}
    \hat{\mathbf{Y}}_i= \mathbf{A} \hat{\mathbf{X}}_i +\hat{\mathbf{Z}}_i,~i\in \{1,\cdots,
    B\},
\end{equation}
which implies that the input symbols on the same row of $\mathbf{X}$
experience the same fading gain, i.e., they are transmitted over the
same block. Since $\hat{\mathbf{Z}}_i$'s are i.i.d random vectors, we
can view this channel as $K$ independent parallel channels with each
channel corresponding to a block. Hence, $KB$ uses of the original
channel corresponds to $B$ uses of the $K$ parallel channels in
(\ref{eq: parallel ch model}). The parallel channels over which a
codeword is transmitted are determined by the channel state $h_1, h_2,
\cdots, h_K$, which can also be viewed as a composite channel
\cite{Berry00}\cite{JWolf}\footnote{A composite channel is a compound
channel with prior probabilities.} that consists of a family of
channels $\{\Gamma(\theta), \theta\in\Theta \}$ indexed by a particular
set of $\Theta$. For a block fading channel with delay constraint $K$,
it can be modeled as a composite channel $\{\Gamma(\theta): \theta \in
\Theta_K\}$ as follows: Let $\Theta_{K} \subset \mathbb{R}^{K}$ be the
set of all length-$K$ sequences of channel gains
$\mv{\theta}_K=\{h_1,h_2,\cdots,h_K\}$, which occurs with probability
$\pi_\theta$ under the joint distribution of $\{H_1,H_2,\cdots,H_K\}$.
For each $\mv{\theta}_K=\{h_1,h_2,\cdots,h_K\}\in \Theta_K$, we
associate a channel $\Gamma(\mv{\theta}_K)$, where
$\Gamma(\mv{\theta}_K)$ consists of $K$ parallel Gaussian channels. Let
$\mv{\alpha}_K=\{\alpha_{1},\alpha_{2},\cdots\,\alpha_{K}\}$ be the
fading power gain vector, i.e., $\alpha_i=|h_i|^2, i=1,\ldots,K$, and
$\mv{P}_K=\{P_{1},P_{2},\cdots,P_{K}\}$ be the transmit power
allocation vector. For a given set of $\mv{\theta}_K$ and $\mv{P}_K$,
the maximum average mutual information rate over channel
$\Gamma(\mv{\theta}_K)$ is \cite{Berry00}:
\begin{equation} \label{eq:BF channel capacity}
C_{\rm BF}(\mv{\theta}_K,\mv{P}_K, K) =
\frac{1}{K}\sum_{i=1}^{K}\log(1+\alpha_{i}P_{i}),
\end{equation}
where we assume a unit noise variance throughout this paper.

In our model of the dying channel, the delay constraint is random
rather than deterministically equal to $K$ due to the fact that a
random attack may happen within any block out of the $K$ blocks or may
not happen at all within the $K$ blocks. If the fatal attack happens
during the transmission, the current block and the blocks after the
attack moment will be discarded. An outage occurs whenever the total
mutual information of the surviving blocks normalized by $K$ is less
than the transmitted code rate. Therefore, the dying channel is
non-ergodic and an appropriately defined outage capacity serves as the
reasonable performance measure.

Let $T$ be the random attack time that is normalized by the block
length. As we know from the results of parallel Gaussian channels
\cite{TCover}, with random coding schemes, we can decode the codeword
even if the attack happens within the $K$ blocks as long as the average
mutual information of surviving blocks is greater than the code rate
$R$ of the transmission, i.e., if we have
\begin{equation*}\label{eq:dying channel rate}
\frac{1}{K}\sum_{i=1}^{L}\log(1+\alpha_{i}P_{i}) \geq R,
\end{equation*}
then the codeword is decodable, where the random integer $L=\min(K,
\lfloor T \rfloor)$ with $\lfloor \cdot \rfloor$ being the flooring
operator.

Hence the outage capacity of a dying channel can be formally defined as
follows:
\begin{definition}
The outage capacity of a $K$-block BF-AWGN dying channel with an
average transmit power constraint $P$ and a required outage probability
$\eta$ is expressed as
\begin{eqnarray}\label{eq:dying channel outage capacity}
&C_{\rm out}(P,\eta)& = \max_{K}\sup_{\mv{P}_K: \sum_{i=1}^K P_k\leq
KP} \bigg\{R: \nonumber \\ &&
\Pr\{\frac{1}{K}\sum_{i=1}^{L}\log(1+\alpha_{i}P_{i})<R \}<\eta
\bigg\}.
\end{eqnarray}
\end{definition}
Note that the outage probability above is defined over the
distributions of the $\alpha_i$'s and $T$, where we assume that the
$\alpha_i$'s and $T$ are independent of each other and the transmitter
does not know the values of the $\alpha_i$'s and $T$ \emph{a priori},
but knows their distributions. As we see from (\ref{eq:dying channel
outage capacity}), there are two sets of variables to be optimized: One
is the number of coding blocks $K$, and the other is the power
allocation vector $\mv{P}_K$. From the perspective of optimal
transmission schemes, the outage capacity maximization problem is
equivalent to the outage probability minimization problem
\cite{Caire99}. In the next section, we first study the optimal coding
length $K$ to ``match'' the attack time in a probabilistic sense such
that the outage probability is minimized.

\section{Optimal Coding Length with Uniform Power Allocation}\label{sec:uniform power}
As discussed before, we can optimize over the coding length $K$ and the
power vector $\mv{P}_K$ to achieve the maximum outage capacity (or
equivalently the minimum outage probability). If a uniform power
allocation strategy is adopted, the only thing left for optimization is
the coding length $K$. On one hand, we can have a larger $L=\min(K,
\lfloor T \rfloor)$ by increasing $K$, meaning that we potentially have
higher diversity to achieve a lower outage probability. On the other
hand, a larger $K$ incurs a higher percentage of blocks being lost
after the attack such that the average achievable mutual information
per block is lower, and hence results in a larger outage probability.
Since the random attack determines the number of surviving blocks and
$K$ determines the average base, we are interested in finding a proper
value of $K$ to ``match'' the random attack property in the sense that
the outage probability is minimized.

With uniform power allocation, according to the law of total
probability, the outage probability can be rewritten as a summation of
the probabilities conditioned on different numbers of surviving blocks,
i.e.:
\begin{eqnarray}\label{eq:expand outage uniform p}
 &  & \Pr\left\{\frac{1}{K}\sum_{i=1}^{L}\log(1+\alpha_{i} P)<R\right\}\nonumber \\
 & = & w_{0}+\Pr\{A_1\}w_{1}+\Pr\{A_2\}w_{2}+\cdots\nonumber \\
 &  & +\Pr\{A_{K-1}\}w_{K-1}+\Pr\{A_K\}w_{K}^{*},
\end{eqnarray}
where $A_j=\{\sum_{i=1}^{j}\log(1+\alpha_i P)<KR\} ~\textrm{for}~
j=1,\cdots,K $, $w_{i}=\Pr(i<T \leq i+1) ~\textrm{for}~ i=0,\cdots,K-1
$, and $w_{K}^{*}=\Pr(T>K)$. Given the distributions of $\alpha_i$ and
$T$, in general, there are no tractable closed-form expressions for
$\Pr\{A_j\}$'s. Alternatively, we could first seek the bounds of the
outage probability and then study more  exact forms for some special
cases where we show how to find the optimal $K$.
\subsection{Outage Probability Lower Bound}\label{subsec:non-uniform fading lower bnd}
Notice that the following relationship holds:
\begin{eqnarray}\label{eq:inclusion lower}
    & & \Pr\left\{\sum_{i=1}^{j}\log(1+\alpha_{i}P)<KR \right\} \nonumber\\
   &\geq & \prod_{i=1}^{j}\Pr\left\{\log\left(1+\alpha_i
   P\right)<\frac{KR}{j}\right\}.\nonumber
\end{eqnarray}
Since the fading gains $\alpha_i$'s of different blocks are i.i.d, we
have
\begin{eqnarray}
    & &\prod_{i=1}^{j}\Pr\left\{\log\left(1+\alpha_i
   P\right)<\frac{KR}{j}\right\}\nonumber\\
    &=& \left\{F\left(\frac{e^{\frac{KR}{j}}-1}{P}\right)\right\}^j, \label{eq:outage lower bound}
\end{eqnarray}
where $F(x)$ is the cumulative distribution function (CDF) of the
random variable $\alpha_i$.

Therefore, with the relationship in (\ref{eq:outage lower bound}), we
have a lower bound for the outage probability in (\ref{eq:expand outage
uniform p}) as
\begin{eqnarray}\label{eq:lower bound}
   & & \Pr\left\{\frac{1}{K}\sum_{i=1}^{L}\log(1+\alpha_i P)<R\right\} \nonumber\\
   &\geq& w_0+\sum_{i=1}^{K-1}\left\{F\left(\frac{e^{KR/i}-1}{P}\right)\right\}^{i}w_i \nonumber\\
   & & +\left\{F\left(\frac{e^{R}-1}{P}\right)\right\}^{K}w_{K}^{*}.
\end{eqnarray}
\subsection{Outage Probability Upper Bound}\label{subsec:non-uniform fading upper bnd}
On the other hand, there exists a simple upper bound for the outage
probability:
\begin{equation*}\label{inclusion upper}
    \Pr\left\{\frac{1}{K}\sum_{i=1}^{j}\log(1+\alpha_{i}P)<R \right\}\leq \prod_{i=1}^{j}\Pr\left\{\log(1+\alpha_i
    P)<KR \right\},~j=1,\cdots,K,
\end{equation*}
hence yielding
\begin{equation*}\label{upper bnd given T}
    \Pr \left\{\frac{1}{K}\sum_{i=1}^{j}\log(1+\alpha_{i}P)<R  \right\} \leq
    \left\{F\left(\frac{e^{KR}-1}{P}\right)\right\}^j,
    ~j=1,\cdots,K.
\end{equation*}
Therefore, an upper bound for the outage probability in (\ref{eq:expand
outage uniform p}) is given as
\begin{eqnarray}\label{eq:upper bound}
    & & \Pr\left\{\frac{1}{K}\sum_{i=1}^{L}\log(1+\alpha_i P)<R\right\} \nonumber\\
   &\leq& w_0+\sum_{i=1}^{K-1}\left\{F\left(\frac{e^{KR}-1}{P}\right)\right\}^i w_i \nonumber\\
   & & + \left\{F\left(\frac{e^{KR}-1}{P}\right)\right\}^{K}w_{K}^{*}.
\end{eqnarray}

\subsection{High SNR Rayleigh Fading Case}\label{sub: high snr rayleigh}
From the previous discussion, we know how to bound the outage
probability in terms of $K$ with the general SNR values. However, there
usually exists a significant gap between the lower and upper bounds.
Fortunately, with appropriate approximations in the high SNR regime
\footnote{Here by high SNR, we mean that $P$ is large, i.e., $P\gg 1$.}
for Rayleigh fading, we can obtain a tractable expression for the
outage probability and hence further derive a closed-form solution for
the optimal $K$.

For our $K$-block fading channel model with high SNR values, outage
typically occurs when each sub-channel cannot support an evenly-divided
rate budget (see Exercise 5.18 in \cite{Tse tradeoff}). Thus,
conditioned on the attack time $T$, the outage probability can be
written as :
\begin{eqnarray}\label{eq:outage approx hihg snr}
    p_{out|T}& = & \Pr\left\{\frac{1}{K}\sum_{i=1}^{L}\log(1+\alpha_i P)< R \right\} \nonumber\\
   &\approx& \left(\Pr\left\{\log(1+\alpha_i P)< \frac{K}{L}R
   \right\}\right)^L.
\end{eqnarray}
For Rayleigh fading, we have $\Pr(\alpha_i<1/x)\approx 1/x$ when $x$ is
large. Thus, when SNR is high, we can simplify (\ref{eq:outage approx
hihg snr}) as
\begin{equation}\label{eq:outage given T}
    p_{out|T}\approx \frac{e^{KR}}{P^{L}}.
\end{equation}
With the conditional outage probability given by (\ref{eq:outage given
T}), the overall outage probability is
\begin{eqnarray}\label{eq:outage K}
  p_{out}(K) &=& w_0+\sum_{i=1}^{K}p_{out|T}\cdot p(L=i) \nonumber \\
   &=&w_0+ \sum_{i=1}^{K-1}\frac{e^{KR}}{P^{i}}w_i+\frac{e^{KR}}{P^{K}}w_{K}^{*}.
\end{eqnarray}
Let $G(t)$ be the CDF of the attack time, which is assumed to be
exponentially distributed with parameter $\lambda$. Let
$w_{K}^{*}=1-G(K)$, $w_{i}=G(i+1)-G(i)=e^{-\lambda
i}(1-e^{-\lambda})=\beta^i c$ (for $\forall i < K$) with
$c=1-e^{-\lambda}$ and $\beta=e^{-\lambda}$. We can rewrite
(\ref{eq:outage K}) as
\begin{eqnarray}\label{eq:outage with exp att}
    p_{out}(K)&=&e^{KR}\sum_{i=1}^{K-1}\frac{\beta^i c}{P^i}+\frac{e^{KR}}{P^{K}}[1-G(K)]+w_0\nonumber\\
              &=&e^{KR}c\frac{\frac{\beta}{P}-(\frac{\beta}{P})^{K}}{1-\frac{\beta}{P}}+\frac{1-G(K)}{P^{K}e^{-KR}}+w_0.
\end{eqnarray}
For high SNR, with $0<\beta <1$, $\frac{\beta}{P}$ is small. Hence,
$\frac{\beta/P-(\beta/P)^{K}}{1-\beta/P}\approx
\frac{\beta/P}{1-\beta/P}$ when $K\geq 2$, and (\ref{eq:outage with exp
att}) can be approximated to:
\begin{equation}\label{eq:outage high SNR}
  p_{out}(K) \approx  \xi e^{KR}+\frac{1}{P^{K}e^{(\lambda-R)K}}+w_0,
\end{equation}
where $\xi = (1-e^{-\lambda})\frac{\beta/P}{1-\beta/P}$. In order to
obtain the optimal $K$ by minimizing $p_{out}(K)$, we first treat
(\ref{eq:outage high SNR}) as a continuous function of $K$, although
$K$ is an integer.

Let us first consider the convexity of (\ref{eq:outage high SNR})
over a real-valued $K$. By taking the second-order derivative of
(\ref{eq:outage high SNR}) over $K$, we have the following:
\begin{equation}\label{eq:sec order}
  \frac{\partial ^2 p_{out}(K)}{\partial K^2} = \xi R^2 e^{KR}+\frac{[\lambda+\log P-R]^2}{(P e^{\lambda-R})^K}.
\end{equation}
Since we have $\lambda > 0$ and $1-\beta/P >0$ in the high SNR regime,
it holds that $\xi>0$. Therefore,  (\ref{eq:sec order}) is non-negative
in the high SNR regime, which means that (\ref{eq:outage high SNR}) is
convex over real-valued $K$.

Given the convexity of (\ref{eq:outage high SNR}), the optimal $K$ can
be derived by setting its first-order derivative to zero and finding
the root. Consequently, the optimal solution $K^{*}$ is obtained as
follows:
\begin{equation*}\label{opt K}
    K^{*}=\log\left[\frac{\lambda+\log P-R}{\xi R}\right]\frac{1}{\lambda + \log P}.
\end{equation*}
Obviously, $K^{*}$ is unique given a set of $\xi, P, R$, and $
\lambda$. Since a feasible $K$ for the original problem should be an
integer, we need to choose the optimal integer solution from $\lfloor
K^{*} \rfloor$ and $\lceil K^{*} \rceil$, whichever gives a smaller
value of (\ref{eq:outage high SNR}).
\begin{figure}[htbp]
 \centering
   \includegraphics[width=0.6\textwidth]{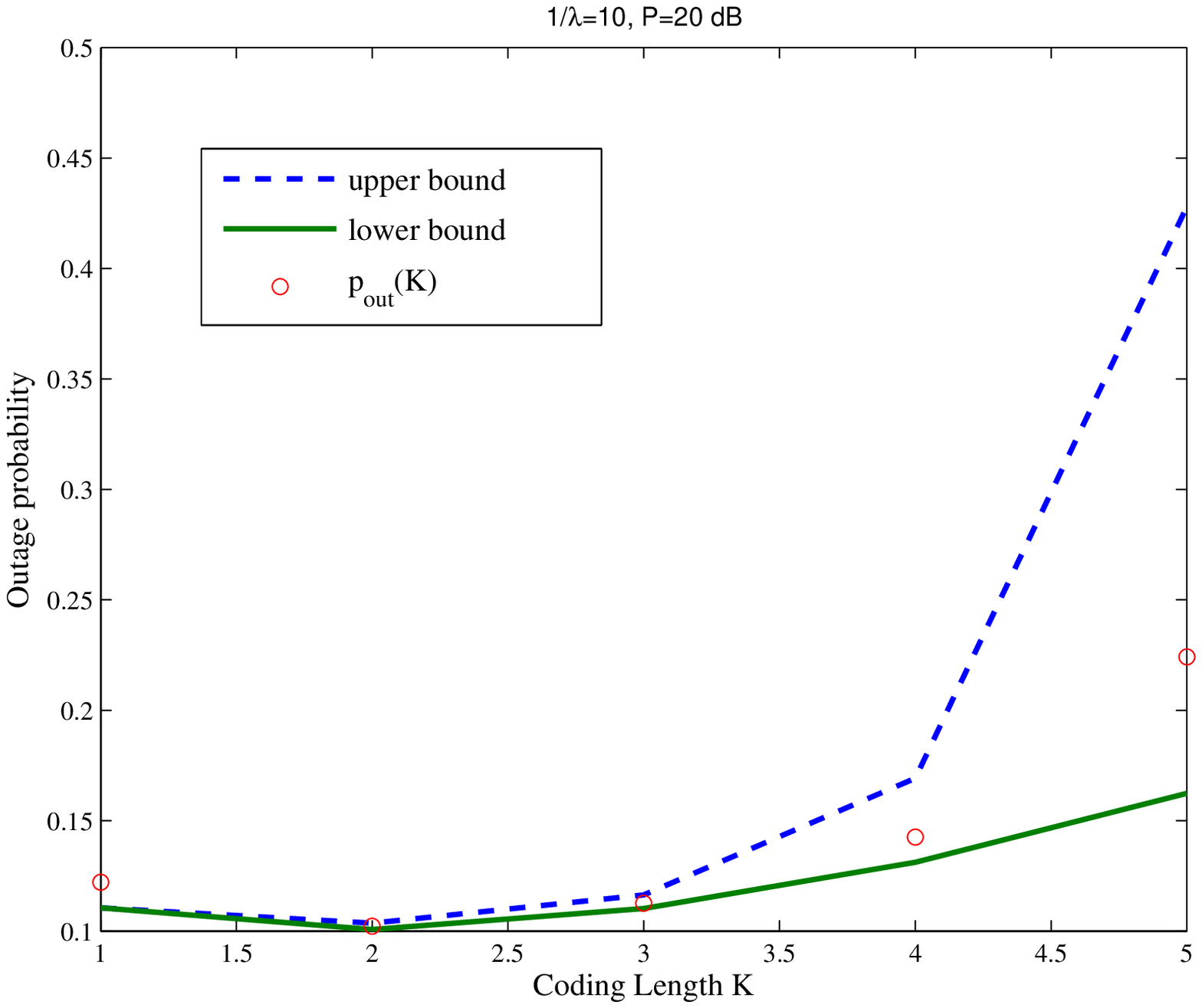}\\
   \caption{Outage probability vs. coding length $K$, P=20dB.}\label{fig:outage vs K 20db}
\end{figure}

\begin{figure}[htbp]
\begin{center}
  \includegraphics[width=0.6\textwidth]{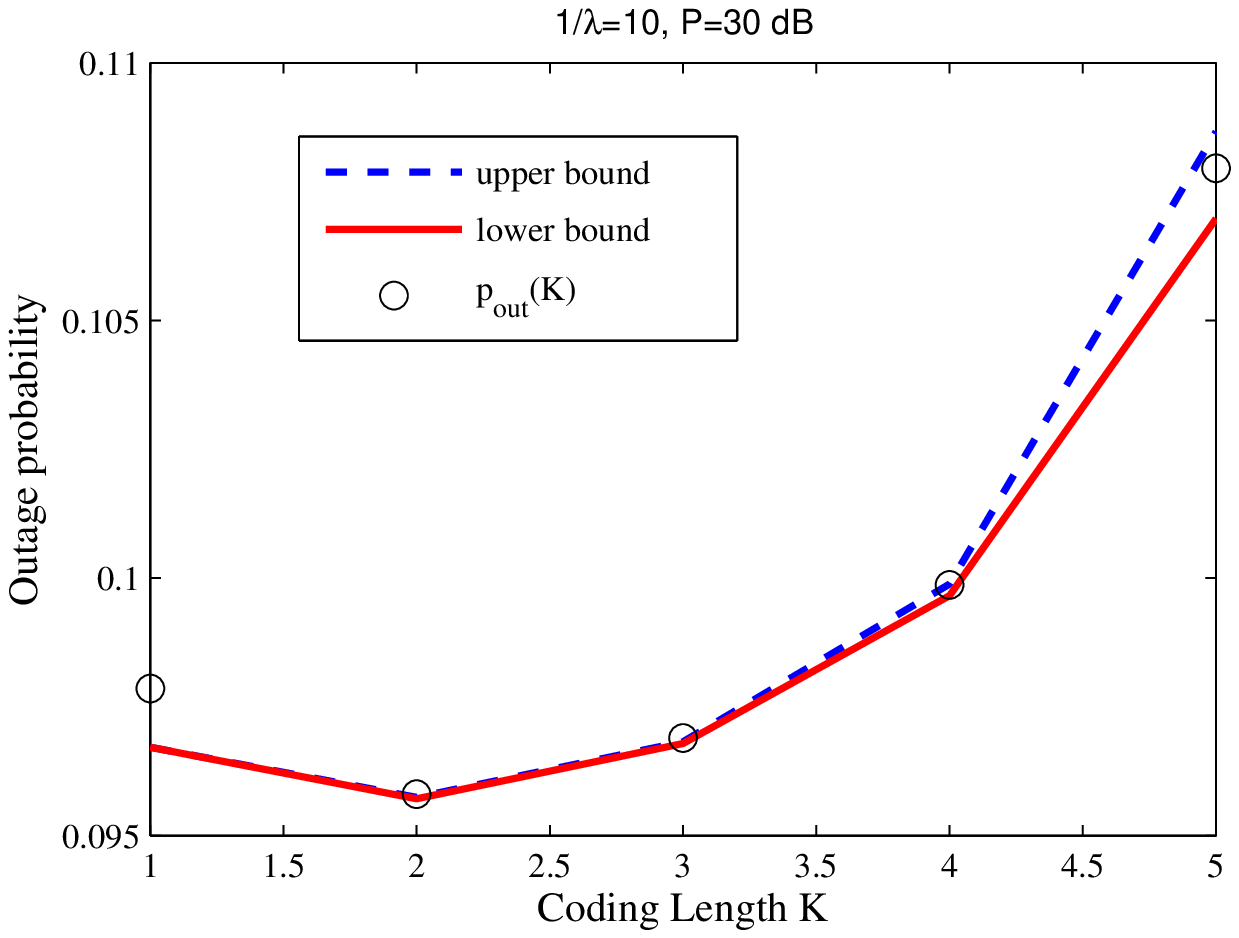}\\
     \caption{Outage probability vs. coding length $K$, P=30dB.}\label{fig:outage vs K 30db}
  \end{center}
\end{figure}
In Fig. \ref{fig:outage vs K 20db} and Fig. \ref{fig:outage vs K 30db},
we plot the coding length $K$ versus the outage probability. We assume
that the fading is Rayleigh, the random attack time is exponentially
distributed with parameter $1/\lambda=10$ (normalized by the
transmission block length), the target rate is $R=1$ nats/s/Hz, and the
transmit power is set as $20$ dB and $30$ dB, respectively. As shown in
Fig. \ref{fig:outage vs K 20db}, the dashed curve and the solid curve
are the lower and upper bounds given by (\ref{eq:lower bound}) and
(\ref{eq:upper bound}), respectively. The circles are obtained by using
(\ref{eq:outage high SNR}). As can be seen, firstly, the high-SNR
approximation in (\ref{eq:outage high SNR}) is quite accurate. The
circles are located between the upper and the lower bounds except for
$K=1$. This is due to the fact that when $K=1$,
$\frac{\beta/P-(\beta/P)^{K}}{1-\beta/P}\approx
\frac{\beta/P}{1-\beta/P}$ does not hold . Secondly, we see that there
exists a minimum outage probability over $K$ as shown in Fig.
\ref{fig:outage vs K 30db}. At last, comparing Fig. \ref{fig:outage vs
K 20db} and Fig. \ref{fig:outage vs K 30db}, we see that the upper and
lower bounds get closer as the SNR increases with the values from
(\ref{eq:outage high SNR}) are in between; hence the approximation in
(\ref{eq:outage high SNR}) becomes more accurate.

\subsection{Low SNR Regime with Arbitrary Fading}\label{sub: low snr}
When SNR is low, we have $\log(1+\alpha_i P) \approx \alpha_i P$. Thus,
when we span a codeword over $K$ blocks, the outage probability
conditioned on $T$ is given as
\begin{eqnarray}\label{eq:outage approx low snr}
  p_{out|T}^{par} &=& \Pr\left\{\sum_{i=1}^{L}\log(1+\alpha_i P)< KR \right\} \nonumber \\
       &\approx & \Pr\left\{\sum_{i=1}^{L}\alpha_i< KR/P\right\}.
\end{eqnarray}
When using a repetition transmission (over blocks), the outage
probability is given as
\begin{eqnarray}\label{eq:outage repetition}
  p_{out|T}^{rep} &=& \Pr\left\{\log(1+\sum_{i=1}^{L}\alpha_i P)< KR \right\} \nonumber\\
       &\approx & \Pr\left\{\sum_{i=1}^{L}\alpha_i< KR/P\right\}.
\end{eqnarray}
Comparing (\ref{eq:outage approx low snr}) and (\ref{eq:outage
repetition}), we see that the outage performances of these two schemes
are the same in the low-SNR regime. This is due to fact that in low SNR
regime it is SNR-limited rather than degree-of-freedom-limited such
that coding over different blocks does not help with decreasing the
outage probability. Hence, repetition transmission is approximately
optimal for a dying channel in the low SNR regime.

\section{Joint Optimization over Coding Length and Power Allocation}\label{sec: joint opt}
In the previous section, we investigated the optimal coding length $K$
that minimizes the outage probability by assuming uniform power
allocation. We now consider optimizing over both the coding length $K$
and the power vector $\mv{P}_{K}$ to minimize the outage probability.
We note that optimizing over $K$ is in general a 1-D search over
integers, which is not complex. Since the main complexity of solving
(\ref{eq:dying channel outage capacity}) lies in the optimization over
$\mv{P}_{K}$, we first focus on the outage probability minimization
problem over $\mv{P}_{K}$ for a given fixed $K$, which is expressed as:
\begin{eqnarray}\label{eq:outage min}
\min_{\mv{P}_K} & & \Pr \left\{\frac{1}{K}\sum_{i=1}^{L}\log(1+\alpha_{i}P_{i})<R \right\} \nonumber \\
\textrm{s.t.} & & \frac{1}{K}\sum_{i=1}^{K}P_{i} \leq P.
\end{eqnarray}
After obtaining the optimal outage probabilities conditioned on a range
of $K$ values, we choose the minimum one as the global optimal value.

\subsection{Properties of Optimal Power Allocation}\label{subsec:opt power profile}
We start solving the above optimization problem by investigating the
general properties of the optimal power allocation over a dying channel
for a given $K$.

Let $E_j$ be the event that
\begin{equation*}\label{event}
    E_j=\left\{\frac{1}{K}\sum_{i=1}^{j}\log(1+\alpha_i
    P_i)<R\right\},~j=1,\cdots,K.
\end{equation*}
It is obvious that the events $E_j$'s are decreasing events, which
means $E_1\supseteq E_2\supseteq \cdots \supseteq E_K$. With the law of
total probability, we can expand the outage probability in the
objective of (\ref{eq:outage min}) as follows,
\begin{eqnarray}
   p_{out}(K)&= & \Pr\left\{\frac{1}{K}\sum_{i=1}^{L}\log(1+\alpha_i P_i)<R\right\} \nonumber\\
   &=& w_0+\Pr\{E_1\}w_1+\Pr\{E_2\}w_2+ \cdots \nonumber\\
   & & +\Pr\{E_{K-1}\}w_{K-1}+\Pr\{E_{K}\}w_{K}^{*}, \label{eq:expand outage}
\end{eqnarray}
where $w_{i}$'s are defined in Section \ref{sec:uniform power}. With
the above result, we then discuss the optimal power allocation for a
dying channel under different conditions.

\subsubsection{Optimal Power Allocation over i.i.d. Fading}\label{subsubsec:opt profile}
\begin{thm}\label{thm:i.i.d fading no-increasing}
When fading gains over blocks are i.i.d., the optimal power allocation
profile is non-increasing.
\end{thm}
\begin{proof}
The proof is provided in Appendix\ref{app: thm1}.
\end{proof}
This is a general result regardless of the specific distributions of
fading gains. That is, the optimal power vector lies in a convex cone
$\mathcal{D}_{+}=\{\mv{P}_{K}\in \mathbb{R}_{+}^{K}: P_1 \geq P_2 \geq
\cdots \geq P_K\}$, no matter what distribution the fading gain
follows, as long as the i.i.d. assumption holds.

\subsubsection{Optimal Power Allocation over Identical Fading Gains}\label{subsubsec:identical fading}
Now we consider the case where the fading gains over all the blocks are
the same, while they are still random. This represents the case where
fading gains are highly correlated in time.
\begin{thm}\label{thm:id fading K is 1}
When the fading gains $\alpha_i$'s are the same, the optimal coding
length is $K=1$ with $P_1=P$.
\end{thm}
\begin{proof}
The proof is provided in Appendix\ref{app: thm2}.
\end{proof}
This assertion implies that the optimal transmission scheme for a
highly correlated dying channel is to simply transmit independent
blocks instead of jointly-coded blocks.

\subsection{Power Allocations for Some Special Cases}
When the fading gain falls into some special distributions, we can
further convert the corresponding optimization problem into convex ones
and derive the optimal power vector efficiently.
\subsubsection{Optimal Power Allocation over i.i.d. Rayleigh Fading in High SNR Regime}
Given (\ref{eq:outage approx hihg snr}) and conditioned on the attack
time $T$, the conditional outage probability can be written as:
\begin{eqnarray}\label{eq:outage approx high snr i.i.d}
    p_{out|T}& = & \Pr\left\{\sum_{i=1}^{L}\log(1+\alpha_i P_i)< KR \right\} \nonumber\\
   &\approx& \prod_{i=1}^{L}\Pr\left\{\log(1+\alpha_i P_i)< \frac{K}{L}R
   \right\}.
\end{eqnarray}
For Rayleigh fading, we have $\Pr(\alpha_i<1/x)\approx 1/x$ when $x$ is
large. Thus, when SNR is high, we can simplify (\ref{eq:outage approx
high snr i.i.d}) as
\begin{equation}\label{eq:outage given T i.i.d}
    p_{out|T}\approx \frac{(e^{KR/L}-1)^{L}}{\prod_{i=1}^{L}P_{i}}.
\end{equation}

The outage probability with Rayleigh fading in high SNR is approximated
as below by substituting (\ref{eq:outage given T i.i.d}) into
(\ref{eq:expand outage}):
\begin{equation}\label{eq:outage_pi}
  p_{out}(K) \approx  w_{0}+\frac{e^{KR}-1}{P_{1}}w_{1}+
  \frac{(e^{KR/2}-1)^2}{P_{1}P_{2}}w_{2}+\cdots+\frac{(e^{KR/K}-1)^K}{\prod_{i=1}^{K}P_{i}}w_{K}^{*}.
\end{equation}
Denoting $c_{i}=w_{i} (e^{KR/i}-1)^i $, we further simplify
(\ref{eq:outage_pi}) as
\begin{equation*}\label{eq:outage pi simple}
 p_{out}(K) \approx  w_{0}+\frac{c_1}{P_{1}}+
  \frac{c_2}{P_{1}P_{2}}+\cdots+\frac{c_{K}}{\prod_{i=1}^{K}P_{i}}.
\end{equation*}

Since the optimal power vector lies in a convex cone as shown in
Theorem \ref{thm:i.i.d fading no-increasing},  the problem can be
formulated as a convex optimization problem (refer to Appendix\ref{app:
cvx high snr} for the convexity proof):
\begin{eqnarray}\label{eq: cvx high snr}
 \mathop {\min }\limits_{\mv{P}_K \in D_{+}} & & w_{0}+\frac{c_1}{P_{1}}+\frac{c_2}{P_{1}P_{2}}+\cdots+\frac{c_{K}}{\prod_{i=1}^{K}P_{i}} \nonumber \\
  \textrm{s.t.} & & \sum_{i=1}^{K} P_{i}\leq KP,
\end{eqnarray}
where $D_{+}=\{\mv{P}\in \mathbb{R}_{+}^{K}: P_{1}\geq P_{2}\geq \cdots
\geq P_{K}\geq 0\}$ is a convex cone. Thus, the optimal power vector
can be efficiently solved with standard convex optimization algorithms
such as the interior point method \cite{Boyd}.

\begin{figure}[htbp]
\begin{center}
  \includegraphics[width=0.5\textwidth]{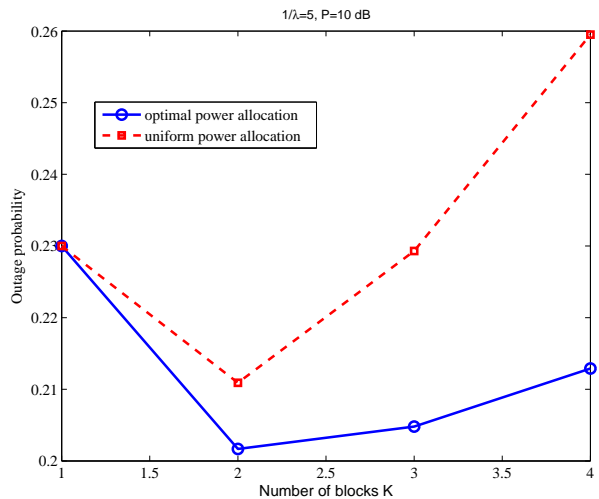}\\
  \caption{Outage probability with non-uniform and uniform power allocation.}\label{fig:cvx vs unifm}
\end{center}
\end{figure}

The simulation results are shown in Fig. \ref{fig:cvx vs unifm}, where
we set the simulation parameters as: $R=0.5$ nats/s/Hz, $1/\lambda=5$
for the exponential random attack, and average power $P=10 $ dB. As we
can see, the power vector derived by solving problem (\ref{eq: cvx high
snr}) achieves better performance in terms of the outage probability
than the uniform power allocation case.

\subsubsection{Optimal Power Allocation over i.i.d. Log-normal Fading}
When the fading gain has a log-normal distribution, we can also
approximate the problem as a convex one by minimizing the upper bound
of the objective function. Since we have
\begin{equation*}\label{log upper bound}
  \sum_{i=1}^{L}\log(\alpha_i P_i) < \sum_{i=1}^{L}\log(1+\alpha_i
  P_i),
\end{equation*}
the outage probability is upper-bounded as follows:
\begin{eqnarray}
  \nonumber & & \Pr\{\sum_{i=1}^{L}\log(1+\alpha_{i}P_{i})<KR\} \\
  \nonumber &<& \Pr\{\sum_{i=1}^{L}\log(\alpha_{i}P_{i})<KR\} \\
  \nonumber &=& w_{0}+\sum_{n=1}^{K}\Pr\{\sum_{i=1}^{n}\log\alpha_{i}< KR -\sum_{i=1}^{n}\log P_{i}\}w_{n}.\label{eq:upper bound inequality}
\end{eqnarray}

Thus, the optimization problem of (\ref{eq:outage min}) is translated
into the following problem, where we essentially minimize the upper
bound:
\begin{eqnarray}
   \min_{\mv{P}_K} & & w_{0}+\sum_{n=1}^{K}\Pr\{\sum_{i=1}^{n}\log\alpha_{i}<KR-\sum_{i=1}^{n}\log P_{i}\}w_{n}\nonumber\\
   \textrm{s.t.} & & \sum_{i=1}^{K} P_{i} \leq KP \label{eq:upper bound min}.
\end{eqnarray}

Let the $\alpha_{i}$'s be independent and log-normal random variables,
i.e., $\log \alpha_{i}\sim \mathcal{N}(0,1),~\forall i$. Since the sum
of $n$ standard normal random variables is a Gaussian random variable
with zero mean and variance $n$, we have
\begin{equation}\label{eq:sum gaussian cdf}
    \Pr\{\sum_{i=1}^{n}\log \alpha_{i} \leq x\}=\frac{1}{2} \{1+\textrm{erf} (\frac{x}{\sqrt{2n}} ) \},
\end{equation}
where $\textrm{erf}(x)$ is the error function. Substituting
(\ref{eq:sum gaussian cdf}) into (\ref{eq:upper bound min}) yields the
new objective function $p_{out}$:
\begin{equation}\label{eq:outage explicit form}
    p_{out} = w_{0} +\sum_{n=1}^{K}\frac{1}{2} \{1+\textrm{erf}(\frac{KR-\sum_{i=1}^{n}\log P_{i}}{\sqrt{2n}} ) \}w_{n}.
\end{equation}

In general, (\ref{eq:outage explicit form}) is not a convex function.
However, under some special circumstances as described in
Appendix\ref{app: log-normal opt}, the problem in (\ref{eq:upper bound
min}) with the objective replaced by (\ref{eq:outage explicit form})
can be rewritten as a convex problem, which is given as following:
\begin{eqnarray}\label{eq: log-normal opt}
   \nonumber \min_{\mv{P}_K} & & w_{0}+ \sum_{n=1}^{K}\frac{1}{2} \{1+\textrm{erf}(\frac{KR-\sum_{i=1}^{n}\log P_{i}}{\sqrt{2n}} ) \}w_{n}\\
  \nonumber \textrm{s.t.} & & \sum_{i=1}^{K} P_{i}\leq KP \\
   \nonumber & & KR-\log P_{1}   \leq   0 \\
   \nonumber & & KR-\log P_{1}-\log P_{2}   \leq   0\\
   \nonumber & & \cdots \cdots \\
    & & KR-\sum_{i=1}^{K} \log P_{i}   \leq   0 .\label{eq:final opt expression}
\end{eqnarray}
Therefore, efficient algorithms can be applied to solve the above
problem.

\begin{figure}[htbp]
  \begin{center}
  \includegraphics[width=0.5\textwidth]{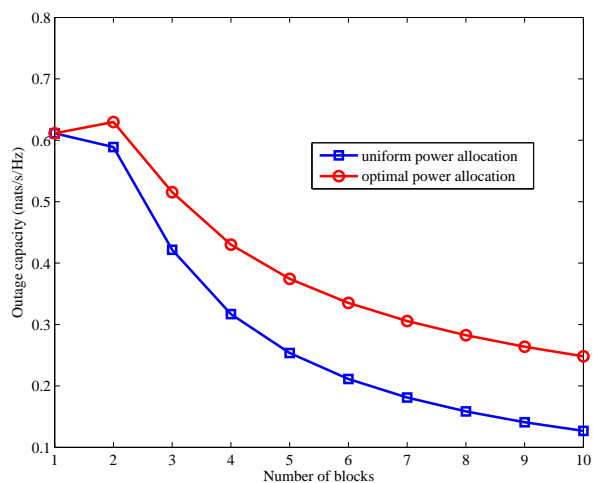}\\
  \caption{Outage capacity v.s. number of blocks $K$, $1/\lambda=4$, average power $P=3$. }\label{fig:outage capacity over K}
  \end{center}
\end{figure}

Numerical results are provided as follows. Assume that the outage
probability target is set as $\eta=0.3$, the attack time is an
exponential random variable with parameter $1/\lambda=4$, and the
fading gains are standard log-normal random variables. As we see from
Fig. \ref{fig:outage capacity over K}, the optimal power allocation
leads to a significantly larger outage capacity over the uniform power
allocation case. Moreover, as $K$ increases, the outage capacity with
the optimal power allocation may even increase to a maximum value while
the outage capacity with uniform power allocation monotonically
decreases. This suggests that, with the potential of a random attack,
we can still span the codeword over more than one block to exploit
diversity and achieve higher outage capacity if the power allocation
and the codeword length $K$ are smartly chosen.

\section{Outage Probability Over Parallel Dying Channels}\label{sec: parallel }
\begin{figure}[htbp]
   \begin{center}
  \includegraphics[width=0.5\textwidth]{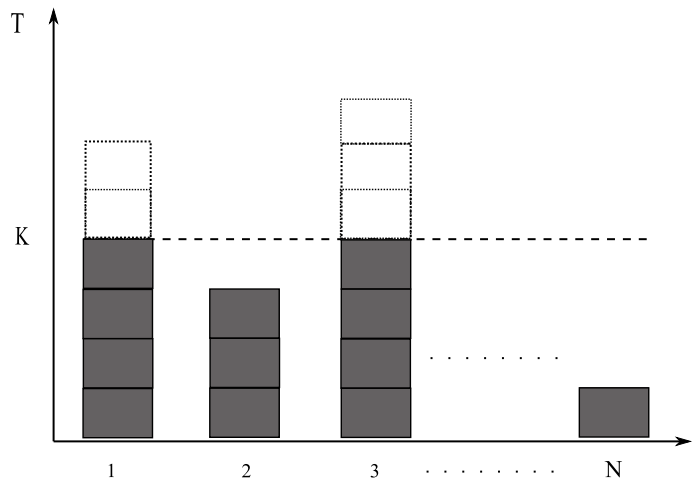}\\
  \caption{Parallel dying channels.}\label{fig: parallel dying channel}
  \end{center}
\end{figure}

In the dying channel example of cognitive radio networks, secondary
users have access to vacant frequency bands that are licensed to
primary users. Some primary users may suddenly show up and take over
some frequency bands, which results in connection losses if these
frequency bands are being used by certain secondary users. Hence, each
sub-channel (a frequency band) may have a different random delay
constraint for information transmission due to the uncertainty of
non-uniform primary user occupancy patterns. Specifically, the above
system can be modeled as follows. Given a link with $N$ parallel
sub-channels as shown in Fig. \ref{fig: parallel dying channel}, the
codeword is spanned in time domain over $K$ blocks and also across all
the $N$ sub-channels. In some sub-channels, random attacks terminate
the transmission before it is completed such that less than $K$ blocks
are delivered. For other sub-channels, $K$ blocks are assumed to be
safely transmitted. What is the maximum rate for reliable communication
over such a link? For the single channel case, it turns out that there
is no way to achieve arbitrarily small outage with a finite transmit
power. However, in this section we show that an arbitrarily small
outage probability is achievable by exploiting the inherent
multi-channel diversity.

In this section, we extend the results of the single dying channel to
the parallel multi-channel case.
\begin{definition}
The outage probability of the parallel multi-channel case is given as
\begin{equation}\label{eq: outage parallel}
    p_{out}(R,P,N)=\Pr\left\{\sum_{i=1}^{N}\frac{1}{K}\sum_{k=1}^{L_i}\log(1+\alpha_{k}^{(i)}
   P/N) <R\right\},
\end{equation}
\end{definition}
where $R$ is the total rate over $N$ sub-channels, $\alpha_{k}^{(i)}$
is the fading gain of block $k$ at sub-channel $i$, $N$ is the number
of sub-channels,  $L_i=\min\{K,\lfloor T_i\rfloor\}$ is the random
number of surviving blocks at sub-channel $i$, $K$ is the number of
blocks over which a codeword is spanned in the time domain, and $P$ is
the total average power such that $P/N$ is the average power for each
sub-channel. Since the asymptotic behavior is concerned, uniform power
allocation is assumed over $N$ sub-channels. According to different
attack models, in the next two sections we investigate the asymptotic
behavior of the above outage probability in two cases: the independent
random attack case and the $m$-dependent random attack case.

\section{Independent Random Attack Case}\label{sec: indp case}
Let the average power $P$ be finite. Since $\log(1+x)\approx x$ if
$|x|\ll 1$, when $N$ is large, we rewrite (\ref{eq: outage parallel})
as
\begin{equation}\label{eq: outage parallel simple}
   p_{out}(R,P,N) \approx \Pr\left\{\frac{1}{N}\sum_{i=1}^{N}\frac{1}{K}\sum_{k=1}^{L_i}\alpha_{k}^{(i)}P<R\right\}.
\end{equation}
We assume that the fading gains $\alpha_{k}^{(i)}$'s are i.i.d., and
let the random variable $Y_i$ be
\begin{equation*}
Y_i=\frac{1}{K}\sum_{k=1}^{L_{i}} \alpha_{k}^{(i)}.
\end{equation*}
For the case of independent random attack, we assume that $L_{i}$'s are
i.i.d., and hence $Y_i$'s are i.i.d..

The outage probability given by (\ref{eq: outage parallel simple}) can
be recast as:
\begin{equation}\label{eq: outage in Y}
  p_{out}(R,P,N) \approx \Pr\left\{\frac{1}{N}\sum_{i=1}^{N} Y_i  < R/P\right\}.
\end{equation}

Since $Y_i$'s are i.i.d., according to the central limit theorem, as
the number of sub-channels $N\rightarrow \infty$, we have
\begin{equation}\label{eq: clt indp}
    \frac{1}{N}\sum_{i=1}^{N} Y_i \rightarrow \mathcal{N} \left(\mu_Y,
    \sigma_Y^2/N\right).
\end{equation}
According to Theorem 7.4 in \cite{Miller04} on the sum of a random
number of random variables, we derive the following relations:
\begin{eqnarray}
  \mu_Y &=&\frac{1}{K}E(L)E(\alpha)\label{eq: mean_Y}\\
  \sigma_Y^2 &=&\frac{1}{K^2}[E(L)Var(\alpha)+Var(L)E(\alpha)^2] \label{eq: var_Y},
\end{eqnarray}
where $\alpha$ is a nominal random variable denoting the fading gain,
$L$ is a nominal integer random variable denoting the number of
surviving blocks of each sub-channel, and $E(\cdot)$ and $Var(\cdot)$
denote the expectation and variance, respectively. As such, the outage
probability can be approximated as:
\begin{equation}\label{eq: apprx outage indp}
    p_{out}(R,P,N) \approx \Phi (\frac{R/P-\mu_Y}{\sigma_Y/\sqrt{N}}).
\end{equation}

As $N\rightarrow\infty$, $\frac{1}{N}\sum_{i=1}^{N} Y_i$ converges to
$\mu_Y$. The outage probability decreases to $0$ over $N$ if $R/P$ is
less than $\mu_Y$, or converges to $1$ if $R/P$ is larger than
$\mu_Y$.\footnote[1]{$R/P$ is interpreted as the rate per unit cost in
\cite{verdu00}. It is interesting to see that the quantity of rate per
unit cost plays an important role here, which is due to the fact that
we operate over both a finite power and a finite coding length.} That
is, even though all sub-channels are subject to fatal attacks, the
outage probability can still be made arbitrarily small when $N$ is
large enough if the rate per unit cost is set in a conservative
fashion, where $\mu_Y$ is a key threshold. This is remarkably different
from the single dying channel case in which the outage probability is
always finite since there are only a finite and random number of blocks
to span a codeword.

\section{$m$-dependent Random Attack Case}\label{sec: m-dep case}
In the previous section, we discussed the case where $L_{i}$'s are
independent. However, in a practical system, such as cognitive radio
networks, the primary users usually occupy a bunch of adjacent
sub-channels instead of picking up sub-channels independently. Thus,
the $L_i$'s across adjacent sub-channels are possibly correlated; and
consequently the achievable rates across adjacent sub-channels are also
correlated. On the other hand, if two sub-channels are far away from
each other, it is reasonable to treat them as independent. Thus, we
assume that $Y_i$'s are strictly stationary \footnote[2]{Call a
sequence $\{X_n, n\geq 1\}$ strictly stationary if, for every $k$, the
joint distribution of $(X_{n+1},\cdots,X_{n+k})$ is independent of
$n$.} and $m$-dependent\footnote[3]{Call a sequence $\{X_n, n\geq 1\}$
$m$-dependent if for any integer $t$, the $\sigma$-fields $\sigma(X_j,
j\leq t)$ and $\sigma(X_j, j\geq t+m+1)$ are independent. Simply put,
$X_{i}$ and $X_{j}$ are independent if $|i-j|>m$.} with the same mean
and variance.

\subsection{Central limit theorem for $m$-dependent
random variables}\label{subsec: CLT m-dp RV}

We first cite the central limit theorem for stationary and
$m$-dependent summands from \cite{DasGupta} (Theorem 9.1 therein).
\begin{thm}[Hoeffding and Robbins]\label{thm: m dep clt}
Suppose $\left\{X_n, n\geq 1\right\}$ is a strictly stationary
$m$-dependent sequence with $E(X_i)=\mu$ and
$Var(X_i)=\sigma^2<\infty$. Then as $N\rightarrow \infty$, we have
\begin{equation}\label{eq: clt mdp }
    \frac{1}{\sqrt{N}}\sum_{i=1}^{N}(X_{i}-\mu) \rightarrow
    \mathcal{N}(0,\upsilon_{m}),
\end{equation}
where
$\upsilon_{m}=\sigma^2+2\sum_{i=1}^{m}\textrm{Cov}(X_{t},X_{t+i})$ with
$\textrm{Cov}(X_{t},X_{t+i})$ the covariance of $X_{t}$ and $X_{t+i}$.
\end{thm}
\begin{proof}
The detailed proof can be found in \cite{Hoeffding48}.
\end{proof}

\subsection{Asymptotic outage probability}\label{subsec:asym outgage}
As assumed, the random sequence $\left\{Y_1, Y_2,\cdots, Y_N\right\}$
is stationary and $m$-dependent, and $Y_i$'s have the same mean and
variance. Then the covariance is given as:
\begin{equation}\label{eq:cov}
    \textrm{Cov}(Y_{i}Y_{i+h})=\bigg\{\begin{array}{cc}
                               0 & |h|>m \\
                               \gamma(h)-\mu_{Y}^{2} & |h|\leq m,
                             \end{array}
\end{equation}
where $\mu_{Y}$ is the expectation of $Y_i$ given in (\ref{eq: mean_Y})
and $\gamma(h)=E(Y_{i}Y_{i+h})$. Meanwhile,
\begin{equation}\label{eq: vm}
    \upsilon_{m}= \sigma_{Y}^2+2\sum_{h=1}^{m}\left(\gamma(h)-\mu_{Y}^{2}\right).
\end{equation}

Due to the fact that the fading gains $\alpha_{p}^{(i)}$ and
$\alpha_{q}^{(i+h)}$ are independent if $p \neq q$ or $h \neq 0$, we
could easily obtain $\gamma(h)$ for $ |h|\leq m, h\neq 0$, as:
\begin{eqnarray}\label{eq: gammaoh}
    \gamma(h) &=& \frac{1}{K^2}E\left[\sum_{p=1}^{L_{i}}\alpha_{p}^{(i)}\sum_{q=1}^{L_{i+h}}\alpha_{q}^{(i+h)}\right] \nonumber\\
    &=& \frac{1}{K^2}E\left[E\left(\sum_{p=1}^{L_{i}}\alpha_{p}^{(i)}\sum_{q=1}^{L_{i+h}}\alpha_{q}^{(i+h)}\bigg|L_{i}L_{i+h} \right)\right] \nonumber\\
    &=& \frac{\mu_{\alpha}^{2}}{K^2}E(L_{i}L_{i+h})
\end{eqnarray}
Assume that $L_i$ and $L_{j}$ have the same correlation coefficient
$\rho$ if $|i-j|\leq m $ and $i\neq j$. The correlation matrix is given
as
\begin{equation*}
C=
   \left(  \begin{array}{*{20}c}
   1 & \rho  &  \cdots  & 0 & 0 & 0  \\
   \rho  & 1 & \rho  &  \cdots  & 0 & 0  \\
    \vdots  & \rho  & 1 & \rho  &  \cdots  & 0  \\
   0 &  \cdots  & \rho  &  \ddots  & \rho  &  \cdots   \\
   0 & 0 &  \cdots  & \rho  &  \ddots  & \rho   \\
   0 & 0 & 0 &  \cdots  & \rho  & 1  \\
\end{array}\right).
\end{equation*}
Note that the following main results can be also derived for other
correlation matrices.

Then  (\ref{eq: gammaoh}) is simplified as
\begin{equation}\label{eq: gammaoh simplified}
    \gamma(h)=\frac{\mu_{\alpha}^{2}}{K^2}(\rho \sigma_{L}^{2}+\mu_{L}^{2}),
\end{equation}
where $\rho$ is a non-negative correlation coefficient, $\mu_L$ and
$\sigma_L$ are the mean and variance of the random variable $L$,
respectively.

Substituting (\ref{eq: mean_Y}), (\ref{eq: var_Y}), and (\ref{eq:
gammaoh simplified}) into (\ref{eq: vm}), we have
\begin{eqnarray}
   \upsilon_{m}&=& \sigma_{Y}^{2}+ 2m\frac{\rho \mu_{\alpha}^{2}\sigma_{L}^{2}}{K^2}\label{eq: vm sigma}\\
   &=& \frac{\mu_{L}\sigma_{\alpha}^{2}}{K^2}+\frac{\mu_{\alpha}^{2}\sigma_{L}^{2}}{K^2}(1+2m\rho)\label{eq: cov}.
\end{eqnarray}

According to Theorem \ref{thm: m dep clt}, we have
\begin{equation}\label{eq: m dep CLT}
    \frac{1}{\sqrt{N}}\sum_{i=1}^{N} \left(Y_{i}-\mu_Y\right) \rightarrow \mathcal{N} (0,
    \upsilon_{m}).\nonumber
\end{equation}
By simple manipulation, we have
 \begin{equation}\label{eq: fin m dep CLT}
    \frac{1}{N}\sum_{i=1}^{N} Y_{i} \rightarrow \mathcal{N} (\mu_{Y},
    \upsilon_{m}/N),
 \end{equation}
where $\mu_{Y}$ is given in (\ref{eq: mean_Y}) and $\upsilon_{m}$ is
given in (\ref{eq: cov}). Hence, the outage probability for the
$m$-dependent random attack case can be approximated as follows when
$N$ is large,
\begin{equation}\label{eq: apprx outage mdp}
    p_{out}(R,P,N) \approx \Phi (\frac{R/P-\mu_Y}{\sqrt{\upsilon_{m}/N}}).
\end{equation}
As we see from (\ref{eq: vm sigma}) that $\upsilon_{m} \geq
\sigma_Y^2$, comparing (\ref{eq: apprx outage indp}) and (\ref{eq:
apprx outage mdp}), we conclude that the outage probability of the
independent attack case is smaller than that of the $m$-dependent case
given the same setting when the rate per unit cost $R/P$ is less than
$\mu_Y$ and the number of sub-channels $N$ is large.

\section{Outage Exponent}\label{sec: outage exponent}
As we learn from the previous sections, the outage probability over
parallel multiple channels goes to zero as $N$ increases if $R/P<\mu_Y$
for both of the two attack cases. In this section, we investigate how
fast the outage probability decreases as $N$ increases for both cases,
which is measured by the outage exponent \cite{widebandAsym08} defined
as
\begin{equation}\label{eq: outage exp definition}
    \mathcal{E}(t)=\mathop {\lim }\limits_{N\rightarrow\infty}\frac{-\log
    p_{out}(R,P,N)}{N},
\end{equation}
where $t=R/P$.

\subsection{Independent Attack Case}\label{subsec: indp case}
According to the results in \cite{widebandAsym08}, we could derive the
outage exponent for the independent attack case as
\begin{equation}\label{eq: outage exp}
  \mathcal{E}(t) = \mathop {\sup }\limits_{s \le 0} \left\{st-\Lambda(s)\right\},
\end{equation}
for $\forall t \leq t_0$, where $t_0=\mu_Y$ and
\begin{eqnarray}\label{eq: Lambda s}
  \Lambda(s) &:=& \log E\left[\exp(s Y_i)\right] \nonumber\\
    &=& \log M_Y(s),
\end{eqnarray}
with $M_{Y}(s)$ the moment generating function of $Y_i$. According to
Theorem 7.5 in \cite{Miller04}, we have $M_Y(s)=h(f(s/K))$ where $h(z)$
and $f(s)$ are the probability generating function of the discrete
random variable $L_i$ and the moment generating function of the
continuous random variable $\alpha_k^{(i)}$, respectively.
\footnote[4]{The moment generating function of the sum of a random
number of random variables, i.e., $S_L=X_1+X_2+\cdots+X_L$, is the
compound function $h(f(s))$, where $L$ is a random integer independent
of $X_i$, $h(z)$ is the probability generating function of $L$, and
$f(s)$ is the moment generating functions of $X_i$.}

\underline{\emph{Example}}: If Rayleigh fading is assumed,
$\alpha_{k}^{(i)}$ is exponentially distributed; hence the
corresponding moment generating function is
$f(s)=(1-s/\lambda_{\alpha})^{-1}$, where $\lambda_{\alpha}$ is the
parameter for the distribution of the $\alpha_{k}^{(i)}$. Assuming that
the random attack time has an exponential distribution, $L$ is an
integer random variable with following distribution:
\begin{eqnarray*}
  w_{0}=\Pr\{L=0\}=\Pr\{0\leq T< 1\}, & & w_{1}=\Pr\{L=1\}=\Pr\{1\leq T<
    2\},\cdots, \\
  w_{K-1}=\Pr\{L=K-1\}=\Pr\{K-1\leq T< K\}, & & w_{K}=\Pr\{L=K\}=\Pr\{K \leq T\}.
\end{eqnarray*}
Thus, we have $h(z)=\sum_{i=0}^{K}w_{i}z^{i}$ and
$M_{Y}(s)=\sum_{i=0}^{K}w_{i}(1-s/(K\lambda_{\alpha}))^{-i}$. Then we
can derive the outage exponent numerically by solving (\ref{eq: outage
exp}) for a given $t$.

\subsection{$m$-dependent Attack Case}\label{subsec: m-dep case} For the
$m$-dependent attack case, the techniques used in deriving the outage
exponent for the independent attack case does not apply any more since
here $Y_i$'s are not independent. In this case, since the outage
probability has an approximate normal distribution, we have
\begin{eqnarray}
  p_{out}(R,P,N)&\approx & \Phi \left(\frac{R/P-\mu_Y}{\sqrt{\upsilon(m)/N}}\right) \nonumber\\
            &= &  Q\left(\frac{\mu_Y-R/P}{\sqrt{\upsilon(m)/N}}\right) \nonumber\\
            & \leq & \exp\left(\frac{{-\left(\frac{\mu_Y-R/P}{\sqrt{\upsilon(m)/N}}\right)^2}}{2}\right)\nonumber\\
            &= &\exp\left(-N\frac{(\mu_Y-R/P)^2}{2\upsilon_{m}}\right)
\end{eqnarray}

Therefore, an approximate outage exponent can be quantified from the
upper bound as
\begin{equation}\label{eq: exponent dep}
    \mathcal{E}_{mdp}(R/P) \approx \frac{(\mu_Y-R/P)^2}{2\upsilon_{m}},
\end{equation}
where $\upsilon_{m}$ is given in (\ref{eq: cov}). The outage exponent
obtained by (\ref{eq: outage exp}) is derived by using the large
deviation techniques. Thus, it is exact while the outage exponent given
by (\ref{eq: exponent dep}) for the $m$-dependent attack case is
approximate. However, when $R/P \ll \mu_Y$, this approximation is
accurate since the exponential bound is tight for the $Q$-function when
its argument is large.

\begin{figure}[htbp]
    \begin{center}
  \includegraphics[width=0.5\textwidth]{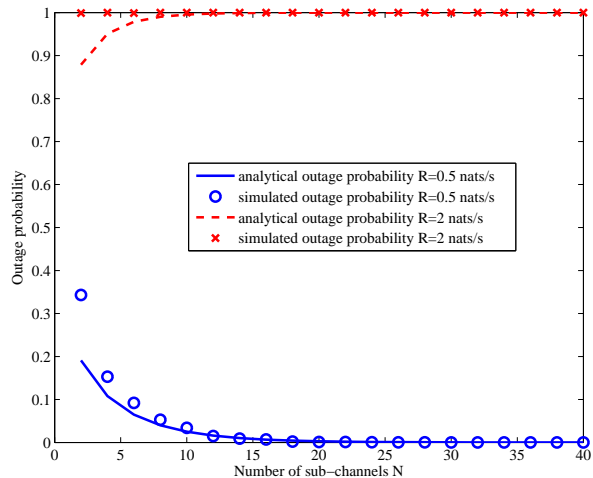}\\
  \caption{Outage probability convergence behavior of the independent case: $\mu_Y$=0.571 and P=2.}\label{fig: convergence indp}
    \end{center}
\end{figure}

Numerical results are provided here to validate our analysis for the
parallel multi-channel case. We choose the random attack time $T$ to be
exponentially distributed with parameter $1/\lambda=5$ and $K$ is
chosen to be 5. Rayleigh fading is assumed and the fading gain
$\alpha_{k}^{(i)}$ is exponentially distributed with parameter 1 and
the noise has unit power. First, we demonstrate the convergence of the
outage probability for the independent attack case and the
$m$-dependent attack case, where the value of $\mu_Y$ according to the
above simulation setup is 0.571. For the independent attack case, as
shown in Fig. \ref{fig: convergence indp}, the solid and dashed curves
are derived by (\ref{eq: apprx outage indp}) while the circles and
crosses are obtained by simulations. We also observe similar
convergence for the $m$-dependent attack case in Fig. \ref{fig:
convergence dep}. In both figures,  the outage probability goes to 0 if
$R/P< \mu_Y$, or goes to 1 if $R/P>\mu_Y$. We see that the accuracy of
Gaussian approximations is acceptable with reasonably large $N$ values.
\begin{figure}[htbp]
    \begin{center}
  \includegraphics[width=0.5\textwidth]{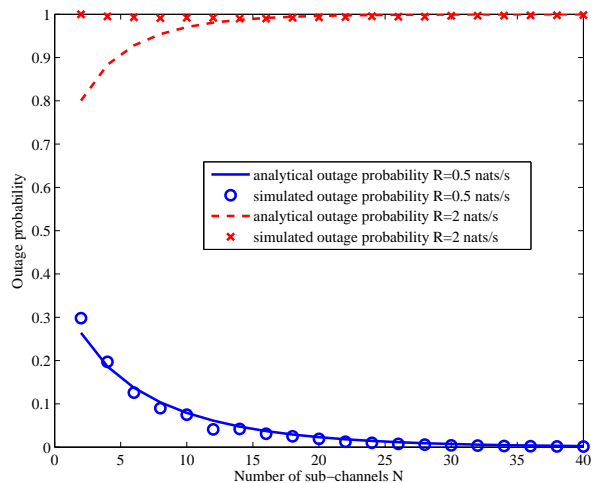}\\
  \caption{Outage probability convergence behavior of the $m$-dependent case: $\mu_Y$=0.571, m=1, $\rho$=0.8, and P=2.}\label{fig: convergence dep}
    \end{center}
\end{figure}

\begin{figure}[htbp]
    \begin{center}
  \includegraphics[width=0.5\textwidth]{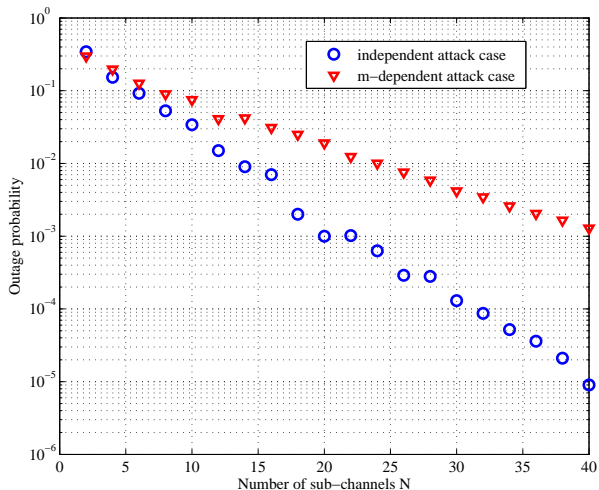}\\
  \caption{Outage probabilities comparison. P=2, R=0.5 nats/s, m=1, and $\rho$=0.8.}\label{fig: outage dep vs indp}
  \end{center}
\end{figure}

Second, we compare the outage probability performance between the
independent case and the $m$-dependent case. Here $P= 2$ and $R=0.5$
nats/s. As shown in Fig. \ref{fig: outage dep vs indp}, the outage
performance of the $m$-dependent case is worse than that of the
independent case even when $m=1$ and $\rho=0.8$. This is due to the
fact that when $R/P<\mu_Y$, the independent attack case is expected to
have a smaller outage probability as we discussed at the end of Section
\ref{subsec: m-dep case}. However, the outage probability of the
$m$-dependent case still decreases to 0 but at a slower rate as the
number of sub-channels $N$ increases, which is caused by the fact that
the $m$-dependent attack case has a smaller outage exponent.

\begin{figure}[htbp]
    \begin{center}
  \includegraphics[width=0.5\textwidth]{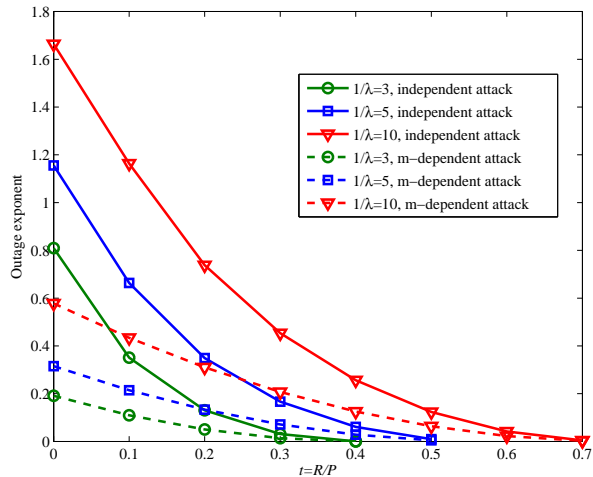}\\
  \caption{Outage exponents for independent and $m$-dependent random attack cases: m=1, $\rho$=0.8, and K=5.}\label{fig: exponent with lambda}
  \end{center}
\end{figure}

In Fig. \ref{fig: exponent with lambda}, we compare the various outage
exponent values between these two cases over the rate per unit cost
$R/P$ with the simulation setup as follows: $K=5$, $m=1$, and
$\rho=0.8$. First, we see that the outage exponent for the independent
attack case is larger than that of the $m$-dependent attack case when
the average attack time $1/\lambda$ is the same. Second, for both of
the independent attack case and the $m$-dependent attack case, a larger
average attack time $1/\lambda$ results in a larger outage exponent.

\section{Conclusion}\label{sec:conclusion}
In this paper, we considered a new type of channels called dying
channels, where a random attack may happen during the transmission. We
first investigated a single dying channel by modeling it as a $K$-block
BF-AWGN channel with a random delay constraint. We obtained the optimal
coding length $K$ that minimizes the outage probability when uniform
power allocation was assumed. Next, we investigated the general
properties of the optimal power allocation for a given $K$. For some
special cases, we cast the optimization problem into convex ones which
can be efficiently solved. As an extension of the single dying channel
result, we investigated the case of parallel dying channels and studied
the asymptotic outage behavior with two different attack models: the
independent-attack case and the $m$-dependent-attack case. It has been
shown that the outage probability diminishes to zero for both cases as
the number of sub-channels increases if the target \emph{rate per unit
cost} is less than a given threshold. Moreover, the outage exponents
for both cases were studied to reveal how fast the outage probability
improves over the number of sub-channels.

\begin{appendices}
\section*{appendix}
\subsection{Proof of Theorem \ref{thm:i.i.d fading
no-increasing}}\label{app: thm1}

Let us consider minimizing the outage probability given by
(\ref{eq:expand outage}). When $K=1$, the proof is trivial.

When $K=2$, the outage probability is
\begin{eqnarray}\label{eq:pout 2}
  p_{out}(2) &=& w_{0}+\Pr\{\log(1+\alpha_1 P_1)<2R\}w_1 \nonumber \\
    & &+ \Pr\{\log(1+\alpha_1 P_1)+\log(1+\alpha_2 P_2)<2R\}w_{2}^{*}. \nonumber
\end{eqnarray}
As we see from the above equation, if $P_1 < P_2$, we have
$\Pr\{\log(1+\alpha_1 P_2)<2R\}<\Pr\{\log(1+\alpha_1 P_1)<2R\}$. Hence,
we can achieve a smaller $p_{out}(2)$ by swapping $P_1$ and $P_2$,
since the last term in $p_{out}(2)$ is not affected by such a swapping
while the second term is decreased.

When $K \geq 3$, for any $j>i,~(i,j\in \{1,\cdots, K\})$, if $P_i <
P_j$, by swapping $P_i$ and $P_j$, all the terms containing both $P_i$
and $P_j$, i.e., all the probability terms in the form of
$\Pr\{\cdots+\log(1+\alpha_i P_i)+\cdots+\log(1+\alpha_j P_j)+\cdots
<KR\}$ will not be affected. However, the probability terms containing
$P_i$ but not $ P_j$ can be decreased by such a swapping. Thus, we
could achieve a smaller outage probability in total.

Therefore, the optimal power allocation profile over i.i.d. fading is
always non-increasing, i.e., $P_1 \geq P_2\geq \cdots \geq P_K \geq 0$.

\subsection{Proof of Theorem \ref{thm:id fading K is 1}}\label{app: thm2}

When the coding length $K=1$, the outage probability is
\begin{equation}
  p_{out}(1)= \Pr\{\log(1+\alpha P)<R\}\Pr\{T>1\}+w_{0}. \nonumber
\end{equation}
When we choose any other arbitrary values for $K$, i.e., $K=M$ and
$M\neq1 $, according to (\ref{eq:expand outage}), the outage
probability is
\begin{eqnarray}\label{eq:pN}
    & &p_{out}(M)\nonumber\\
    &=& \Pr\left\{\frac{1}{M}\sum_{i=1}^{M}\log(1+\alpha P_i)< R\right\}\Pr\{T>M\}\nonumber\\
    & &+w_{0}+\sum_{i=1}^{M-1}\Pr\left\{\frac{1}{M}\sum_{l=1}^{i}\log(1+\alpha P_l)< R\right\}w_{i}.
\end{eqnarray}
Due to the concavity of the $\log$ function, we have
$\frac{1}{M}\sum_{i=1}^{M}\log(1+\alpha P_i)\leq \log(1+\alpha P)$.
Hence,
\begin{equation}\label{eq:pN greater than p1}
     \Pr\left\{\frac{1}{M}\sum_{l=1}^{M}\log(1+\alpha P_l)<R\right\}\geq \Pr\{\log(1+\alpha
 P)<R\}.
\end{equation}
Moreover, it is obvious that summing over only a portion of the $M$
blocks yields an even smaller value, i.e.,
$\frac{1}{M}\sum_{l=1}^{i}\log(1+\alpha P_l) \leq \log(1+\alpha P)$,
with $1\leq i \leq M-1$. If $\exists P_j
>0$, for $i<j\leq M$, the strong inequality holds. Therefore,
we have
\begin{equation}\label{eq:increasig relation of prob}
 \Pr\left\{\frac{1}{M}\sum_{l=1}^{i}\log(1+\alpha P_l)<R\right\} \geq \Pr\{\log(1+\alpha
 P)<R\}.
\end{equation}
Noting that $\sum_{i=1}^{M-1}w_i=\Pr\{1<T\leq M\}$,  and considering
(\ref{eq:pN greater than p1}) and (\ref{eq:increasig relation of
prob}), the following inequality can be derived for (\ref{eq:pN}):
\begin{eqnarray}\label{eq:total pout greater}
 p_{out}(M)& \geq &w_{0}+\Pr\{\log(1+\alpha P)<R\}\cdot \nonumber\\
     & & \left( \Pr\{T>M\}+\Pr\{1 < T \leq M\} \right) \nonumber\\
     &=& p_{out}(1).
\end{eqnarray}
From (\ref{eq:total pout greater}), we see that $p_{out}(1)$ has the
smallest outage probability when fading gains are the same, which means
that the optimal coding length is $K=1$ with $P_1=P$.

\subsection{Convexity of the optimization problem in (\ref{eq: cvx high
snr})}\label{app: cvx high snr}

We first check the Hessian matrix of the objective function in terms of
$P_i$.

\begin{equation}\label{eq:hessian pout}
   \nabla^{2} p_{out}= \nabla^{2}\frac{c_1}{P_{1}}+\cdots+\nabla^{2}\frac{c_K}{\prod_{i=1}^{K}P_{i}}.
\end{equation}
The $j$th term is:

\begin{eqnarray}\label{hessian matrix}
  & & \nabla^{2}\left(\frac{c_{j}}{\prod_{i=1}^{j}P_{i}}\right) \nonumber\\
   \nonumber&=&  c_j \left(
    \begin{array}{ccccc}
    \frac{2}{P_{1}^{3}\prod_{i=2}^{j}P_i} & \frac{1}{P_{1}^{2}P_{2}^{2}\prod_{i=3}^{j}P_i} & \cdots& \frac{1}{P_{1}^{2}P_{j}^{2}\prod_{i=2}^{j-1}P_i} & \mathbf{0} \\
    \frac{1}{P_{1}^{2}P_{2}^{2}\prod_{i=3}^{j}P_i} & \frac{2}{P_{1}P_{2}^{3}\prod_{i=3}^{j}P_i} & \cdots & \frac{1}{P_{2}^{2}P_{j}^{2}\prod_{i=1,i\neq 2}^{j-1}P_i}&  \mathbf{0} \\
    \cdots & \cdots & \ddots & \cdots \\
       & \mathbf{0} &   &   &  0\\
     \end{array}
     \right)
\end{eqnarray}

Let $\mathbf{z} \in \mathbb {R}^K$, then

\begin{equation*}\label{sdp}
    \mathbf{z}^{T}\nabla^{2}\left(\frac{c_{j}}{\prod_{i=1}^{j}P_{i}}\right) \mathbf{z}=
    \frac{1}{\prod_{i=1}^{j}P_{i}}\mathbf{z}^{T}\mv{P}^{(j)}(\mv{P}^{(j)})^{T}\mathbf{z}
    + \mathbf{z}^T \mv{M} \mathbf{z} \geq 0,
\end{equation*}
where $\mv{P}^{(j)}=\left(1/P_1, 1/P_2, \cdots, 1/P_j, 0,\cdots, 0
\right)^T$, and  $\mv{M}= \textrm{diag} \left(\frac{1}{P_{1}^{2}},
\frac{1}{P_{2}^{2}}, \cdots, \frac{1}{P_{j}^{2}}, 0,\cdots, 0\right)$.

Therefore, (\ref{eq:hessian pout}) as the summation of all the $K$
terms is positive semi-definite. Hence $p_{out}$ is a convex function
in terms of $\mv{P}_K$. In addition,  $\mv{P}_K$ lies in a convex cone
as shown in Theorem. \ref{thm:i.i.d fading no-increasing}. Hence the
problem is a convex problem.

\subsection{Sufficient conditions for the convexity of the optimization problem in (\ref{eq: log-normal opt})}\label{app: log-normal opt}
Let $h: \mathbb{R}^{k}\longrightarrow \mathbb{R}$, $g:
\mathbb{R}^{n}\longrightarrow \mathbb{R}^{k}$, and $f=h\circ g :
\mathbb{R}^{n}\longrightarrow \mathbb{R}$ be defined as:
\begin{equation}
    \nonumber f(x)=h(g(x)),~\textbf{dom}~f=\{x\in \textbf{dom}~ g| g(x)\in \textbf{dom}~h\},
\end{equation}
where $\textbf{dom}$ is the domain of a function. The following two
lemmas can be established.
\begin{lemma}
$f$ is convex if $h$ is convex and nondecreasing, and $g$ is convex.
\end{lemma}
\begin{proof}
See Chapter 3 in \cite{Boyd}.
\end{proof}

\begin{lemma}\label{lemma:cvx condition}
The outage probability function given by (\ref{eq:outage explicit
form}) is convex, if
\begin{eqnarray}
  \nonumber KR-\log P_{1} & \leq & 0, \\
  \nonumber KR-\log P_{1}-\log P_{2} & \leq & 0,\\
  \nonumber \cdots \cdots \\
  \nonumber KR-\sum_{i=1}^{K}\log P_{i} & \leq & 0.\label{eq:bunch of constraint}
\end{eqnarray}
\end{lemma}
\begin{proof}
As we know, the error function
$\textrm{erf}(x)=\frac{1}{\sqrt{2\pi}}\int_{0}^{x}e^{-t^2}dt $ is a
convex function for $x \leq 0$ and it is non-decreasing.

Let
\begin{equation*}\label{eq:g of p}
    g_{m}(\mv{P}_K) = \frac{KR-\sum_{i=1}^{m}\log P_{i}}{\sqrt{2m}}.
\end{equation*}
Since $g_{m}(\mv{P}_K)$ is convex and $\textrm{erf}(x)$ is convex for
$x \leq 0$ and nondecreasing, according to Lemma 1, if $g_{m}(\mv{P}_K)
\leq 0$, $\textrm{erf}(g_{m})$ is convex over $\mv{P}_{K}$. Hence, the
objective function given by (\ref{eq:outage explicit form}) is convex
under the conditions given in the lemma.

Since the constraints in (\ref{eq: log-normal opt}) are obviously
convex and the objective function is convex under these constraints,
the problem in (\ref{eq: log-normal opt}) is convex.
\end{proof}
\end{appendices}

\end{document}